\documentclass[twoside]{ae100prg}
\usepackage{graphicx}
\usepackage[breaklinks]{hyperref}
\usepackage{booktabs}
\usepackage{mathptmx} 
\usepackage{fixmath}
\usepackage{bm}

\DeclareSymbolFont{extra}{OML}{cmm}{m}{n}
\DeclareMathSymbol{\varrho}{\mathord}{extra}{'045} 
\DeclareMathSymbol{\nu}{\mathord}{extra}{'027}
\DeclareMathSymbol{\zeta}{\mathord}{extra}{'020}
\DeclareMathSymbol{\kappa}{\mathord}{extra}{'024}
\DeclareMathSymbol{\omega}{\mathord}{extra}{'041}
\DeclareMathSymbol{\Ph}{\mathord}{extra}{'10}
\DeclareMathSymbol{\Omega}{\mathord}{extra}{'12}
\DeclareMathSymbol{\alpha}{\mathord}{extra}{'13}
\DeclareMathSymbol{\beta}{\mathord}{extra}{'14}
\DeclareMathSymbol{\gamma}{\mathord}{extra}{'15}
\DeclareMathSymbol{\delta}{\mathord}{extra}{'16}
\DeclareMathSymbol{\eta}{\mathord}{extra}{'21}
\DeclareMathSymbol{\xi}{\mathord}{extra}{'30}
\DeclareMathSymbol{\varepsilon}{\mathord}{extra}{'42}
\DeclareMathSymbol{\varphi}{\mathord}{extra}{'47}
\DeclareMathSymbol{\pi}{\mathord}{extra}{'31}
\DeclareMathSymbol{\phi}{\mathord}{extra}{'36}
\DeclareMathSymbol{\Phii}{\mathord}{extra}{'10}
\DeclareMathSymbol{\chi}{\mathord}{extra}{'37}
\DeclareMathSymbol{\psi}{\mathord}{extra}{'40}

\newcommand{\dd}{\mathrm{d}}
\newcommand{\ee}{\mathrm{e}}
\newcommand{\ii}{\mathrm{i}}

\newcommand{\eps}{\varepsilon}

\newcommand{\diff}[2]{\frac{\partial#1}{\partial#2}}
\newcommand{\rW}{\varrho}
\newcommand{\zW}{\zeta}

\newcommand{\Ap}{\mathcal A^+}
\newcommand{\An}{\mathcal A^0}
\newcommand{\Am}{\mathcal A^-}

\newcommand{\Ha}{\mathcal{H}^{(1)}}
\newcommand{\Hb}{\mathcal{H}^{(2)}}
\newcommand{\bPhi}{{\mathbold\Phi}}
\newcommand{\Rp}{{\bf R}^{+}}

\begin{document}
\title{Stationary black-hole binaries:\\  A non-existence proof}

\author{Gernot Neugebauer$^1$ and J\"org Hennig$^2$}
\address{$^1$Theoretisch-Physikalisches Institut, Friedrich-Schiller-Universit\"at Jena, Max-Wien-Platz 1, D-07749 Jena, Germany}
\email{neugebauer@tpi.uni-jena.de\\ \mbox{}}
\address{$^2$Department of Mathematics and Statistics, University of Otago, P.O.~Box 56, Dunedin 9054, New Zealand}
\email{jhennig@maths.otago.ac.nz}

\begin{abstract}
We resume former discussions of the question, whether the spin-spin repulsion and the gravitational attraction of two aligned black holes can balance each other. Based on the solution of a boundary problem for disconnected (Killing) horizons and the resulting violation of characteristic black hole properties, we present a non-existence proof for the equilibrium configuration in question. From a mathematical point of view, this result is a further example for the efficiency of the inverse (``scattering'') method in non-linear theories.

\end{abstract}

\section{Introduction}
The examination of time-independent two-body systems dates back to the early days of General Relativity. In a 1922 paper, Rudolf Bach and Hermann Weyl~\cite{Bach1922}
discussed the superposition of two exterior Schwarzschild solutions in Weyl coordinates as a characteristic example for an equilibrium configuration consisting of two ``sphere-like'' bodies at rest. Bach noted that this \emph{static} solution develops a singularity on the portion of the symmetry axis between the two bodies, which violates the elementary flatness on this interval. In a supplement to Bach's contribution, Weyl focused on the interpretation of this type of singularity and used stress components of the energy-momentum tensor to define a \emph{non-gravitational} repulsion between the bodies which compensates the gravitational attraction. Weyl's result is based on some artificial assumptions but implies an interesting question: Are there repulsive effects of \emph{gravitational} origin which could counterbalance the omnipresent mass attraction?

Post-Newtonian approximations tell us that the interaction of the angular momenta of rotating bodies (``spin-spin interaction'') could indeed generate repulsive effects. This is a good  motivation to study, in a rigorous way,  \emph{stationary} two-body problems.

\begin{figure}\label{fig:2BHs}
 \centering
 \includegraphics[scale=0.8]{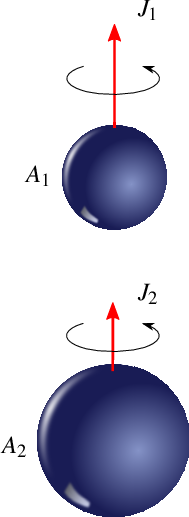}
 \caption{Illustration of two aligned rotating black holes with horizon areas $A_1$, $A_2$ and angular momenta $J_1$, $J_2$.}
\end{figure}

In this contribution we shall summarize the results that we obtained for a stationary two-black-hole system consisting of two aligned rotating  black holes with parallel (or anti-parallel) spins, see Fig.~\ref{fig:2BHs}.
The re\-presentation is based on three recent papers which contain the  \emph{details} of the analysis~\cite{NeugebauerHennig2009,HennigNeugebauer2011,NeugebauerHennig2012}.
The \emph{idea} of our non-existence proof is to construct the exterior gravitational fields of two disconnected Killing horizons, see Fig.~\ref{fig:IntPath} below,
via a boundary problem for the \emph{nonlinear} Ernst equation, which is essentially equivalent to the vacuum Einstein equations. Fortunately, this equation belongs to a class of completely integrable differential equations, which can be mapped to \emph{linear} structures (``Linear Problems''). This fact is the source of powerful solution generating methods such as B\"acklund transformations. It can be shown that a single B\"acklund transformation \cite{Neugebauer1980,Neugebauer1996} applied to Minkowski space creates a Kerr-NUT spacetime which includes the spacetime of the rotating black hole. Since iterative
B\"acklund transformations act as a ``nonlinear superposition principle'', the double-Kerr-NUT solution \cite{KramerNeugebauer1980,Neugebauer1980} was considered to be a good candidate for the solution of the two-horizon problem and extensively discussed in the literature \cite{DietzHoenselaers1985,Hoenselaers1984,HoenselaersDietz1983,Kihara1982,Kramer1986,KramerNeugebauer1980,Krenzer2000,Manko2001,Manko2000,TomimatsuKihara1982,Yamazaki1983}.
However, there was no argument ensuring that this particular solution be the \emph{only} candidate. We have removed this objection and shown that the solution procedure for the \emph{boundary problem} necessarily leads to a subclass of the double-Kerr-NUT solution. Thus we could make use of results derived for the double-Kerr-NUT solution.   
 The result is in line with a theorem of Varzugin~\cite{Varzugin1997,Varzugin1998},
which says that the $2N$-soliton solution by Belinski\u{i} and Zakharov \cite{BelinskiZakharov1978,BelinskiZakharov1979} contains all possible solutions (if any existed) corresponding to an  equilibrium configuration of black holes. The subclass is characterized by a set of restrictions for the parameters of the general double-Kerr-NUT solution. These restrictions, first derived and discussed by Tomimatsu and Kihara \cite{Kihara1982,TomimatsuKihara1982}, ensure the regularity of the double-Kerr-NUT solution on the axis of symmetry and on the horizons. An elegant reformulation of the Tomimatsu-Kihara regularity conditions by Manko, Ruiz and Sanabria-G\'omez~\cite{Manko2000}
made it possible to express black hole quantities such as mass, angular momentum and surface area in terms of  independent parameters. We have made use of these results. Another condition to be satisfied is the positivity of the total mass. Combining the restrictions with symmetry arguments, Hoenselaers and Dietz~\cite{DietzHoenselaers1985,HoenselaersDietz1983} and Krenzer \cite{Krenzer2000}
could show that the double-Kerr-NUT solution cannot describe a configuration consisting of two \emph{identical} black holes. Manko and Ruiz \cite{Manko2001} generalized this result by showing that the regularity conditions imply that at least one of the two horizons has a negative Komar mass. They argued, without giving any explanation, that this peculiarity casts out the double-Kerr-NUT solution. Remarkably, Ansorg and Petroff~\cite{AnsorgPetroff2006}, who described an equilibrium configuration with a positive total mass and a component which has a negative Komar mass, came to an opposite interpretation. However, considerations like these stimulated us to examine further black hole inequalities. Fundamental \emph{local} ``state variables'' of a rotating black hole are its area $A$ and its angular momentum $J$. Indeed, for a single black hole these quantities are restricted by the inequality $8\pi|J|\le A$. Based on results of Ansorg and Pfister~\cite{AnsorgPfister2008}, who examined extremal black holes, 
Hennig, Cederbaum and Ansorg \cite{Hennig2008}, who, following Booth and Fairhurst \cite{BoothFairhurst2008}, studied sub-extremal black holes defined by existence of trapped surfaces (surfaces with a negative expansion of outgoing null rays) in every sufficiently small interior neighbourhood of the event horizon, and Chrus\'ciel, Eckstein, Nguyen and Szybka \cite{Chrusciel2011},
we can assume that each of the two black holes has to satisfy the angular momentum-area inequality \emph{individually}. Surprisingly, Dain and Reiris~\cite{DainReiris2011}
were able to extend its range of application to non-stationary black holes, see also the overview article by Dain~\cite{Dain2011} and references therein.

\section{Mathematical tools}

\subsection{Metric and horizons}
The exterior vacuum gravitational field of axially symmetric and stationary gravitational sources can be described in cylindrical Weyl-Lewis-Papapetrou coordinates $(\rW, \zW, \varphi, t)$\footnote{In the following, we also use the complex coordinates $z=\rW+\ii\zW$ and $\bar z = \rW-\ii\zW$. $t$ is the time coordinate.}, in which the line element takes the form
\begin{equation}
 \dd s^2 = \ee^{-2U}\big[\ee^{2k}(\dd\rW^2+\dd\zW^2)
           +\rW^2\dd\varphi^2\big]
           -\ee^{2U}(\dd t+a\,\dd\varphi)^2,
    \label{LE}
\end{equation}
where the ``Newtonian'' gravitational potential $U$, the gravitomagnetic potential $a$ and the ``superpotential'' $k$ are functions of $\rW$ and $\zW$ alone. At large distances $r=|\sqrt{\rW^2+\zW^2}|\to\infty$ from isolated sources located around the origin of the coordinate system, $r=0$, the spacetime has to be Minkowskian,
\begin{equation}
 r\to\infty:\quad \dd s^2=\dd\rW^2+\dd\zW^2+\rW^2\dd\varphi^2-\dd t^2.
    \label{asflat}
\end{equation}
 Metric (\ref{LE}) admits an Abelian group of motions $G_2$ with the generators (Killing vectors)
\begin{eqnarray}
 &&\xi^i = \delta^i_t,\quad\hspace{1.2mm}\textrm{(stationarity)},\\
 &&\eta^i = \delta^i_\varphi,\quad\textrm{(axisymmetry)},
    \label{stax}
\end{eqnarray}
where the Kronecker symbols $\delta^i_t$, $\delta^i_\varphi$ indicate that $\xi^i$ has only a time $t$-component whereas $\eta^i$ points in the azimuthal $\varphi$-direction. $\eta^i$ has closed compact trajectories about the axis of symmetry and is therefore spacelike off the axis (and the horizons). $\xi^i$ is timelike sufficiently far from the black holes but can become spacelike inside ergoregions.  Obviously,
\begin{equation}
 \ee^{2U} = -\xi^i\xi_i,\qquad a=-\ee^{-2U}\eta_i\xi^i
    \label{mc}
\end{equation}
is a coordinate-free representation of the two relativistic gravitational potentials $U$ and $a$.

 In stationary (and axisymmetric) spacetimes, the \emph{event horizon} as the central black hole property is a \emph{local} 
concept. Consider the Killing vector $\xi'$,
 \begin{equation}
 \xi'=\xi+\Omega\eta
    \label{KV}
\end{equation}
with the norm
\begin{equation}
 \ee^{2V}=-(\xi',\xi')=\ee^{2U}\left[(1+\Omega a)^2-\rW^2\Omega^2\ee^{-4U}\right],
    \label{Vdef}
\end{equation}
where $\Omega$ is a real constant. A connected component of the set of points with $\ee^{2V}=0$, which is a null hypersurface, $(\dd\ee^{2V},\dd\ee^{2V})=0$, is called a \emph{Killing horizon} $\mathcal H(\xi')$,
\begin{equation}
  \mathcal H(\xi'):\quad \ee^{2V}=-(\xi',\xi')=0,\quad (\dd\ee^{2V},\dd\ee^{2V})=0.
    \label{KH} 
\end{equation}
  Since the Lie derivative $\mathcal L_{\xi'}$ of $\ee^{2V}$ vanishes, we have $(\xi',\dd\ee^{2V})=0$. Being null vectors on $\mathcal H(\xi')$, $\xi'$ and $\dd\ee^{2V}$ are proportional to each other,
\begin{equation}
 \mathcal H(\xi'):\quad \dd\ee^{2V}=-2\kappa \xi'.
    \label{kappa}
\end{equation}

 Using the (vacuum) field equations one can show that the \emph{surface gravity} $\kappa$ is a constant on $\mathcal H(\xi')$. $\Omega$ is
the \emph{angular velocity} of the horizon. In black hole thermodynamics, $\kappa$ and $\Omega$ are conjugate to the extensive quantities
$A$ (area) and $J$ (angular momentum), respectively.

\begin{figure}
 \centering
 \includegraphics[width=4.2cm]{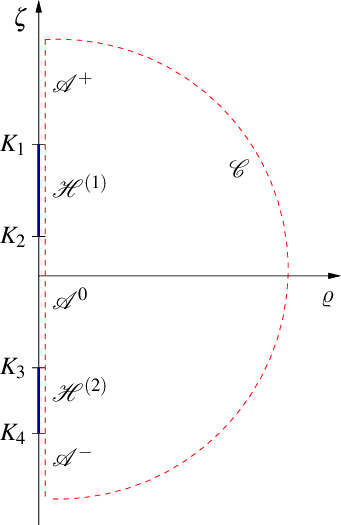}
 \caption{A two-black-hole equilibrium configuration in Weyl-Lewis-Papapetrou coordinates. 
(Adapted from \cite{NeugebauerHennig2009}.)}
 \label{fig:IntPath}
\end{figure}
 In the $\rW$-$\zW$ plane ($t=\textrm{constant}$, $\varphi=\textrm{constant}$) of the Weyl-Lewis-Papapetrou coordinate system (\ref{LE}), the horizons cover a finite portion on the $\zW$-axis ($\rW=0$), see Fig.~\ref{fig:IntPath}, or shrink to a point~\cite{Carter1973}. 
It turns out that extended horizons (``sub-extremal horizons'') and point-like horizons (``degenerate horizons'') require different considerations. Note that a Killing horizon is always a two-surface in the time slice $t=\textrm{constant}$.
The degeneracy to a line or a point is a peculiarity of the special coordinate system.

In this paper, we explain the non-existence proof for extended (sub-extremal) horizons and end up with a brief comment on degenerate horizons.

The dashed line in Fig.~\ref{fig:IntPath} sketches the boundaries of the vacuum region: $\Ap$, $\An$, $\Am$ are the vacuum parts of the $\zW$-axis (axis of symmetry), $\Ha$ and $\Hb$ denote the two Killing horizons, which are located in the intervals $[K_2,K_1]$ and $[K_4,K_3]$ on the $\zW$-axis, and $\mathcal C$ stands for spatial infinity.
The gravitational fields $a$, $k$, $U$ have to satisfy the following boundary conditions
\begin{eqnarray}
 \label{B1}
 \mathcal A^\pm,\An: && \quad a=0,\quad k=0,\\
 \label{B2}
 \mathcal H^{(i)}:       && \quad 1+\Omega_i a=0,\quad i=1,2,\\
 \label{B3}
 \mathcal C:         && \quad U\to 0,\quad a\to 0,\quad k\to 0,         
\end{eqnarray}
where $\Omega_1$ and $\Omega_2$ are the angular velocities of the two horizons. Equations (\ref{B1}) characterize the axis of symmetry (rotation axis). The first relation originates from the second equation in (\ref{mc}), since the compact trajectories of $\eta$ with the standard periodicity $2\pi$ become infinitesimal circles with the consequence $\eta\to 0$. The second relation is a necessary condition for elementary flatness (Lorentzian geometry in the vicinity of the rotation axis). Equation (\ref{B2}) is a reformulation of Eqs.~(\ref{KH}) ($\ee^{2V}=0$) and (\ref{Vdef}) since the horizons are located on the $\zW$-axis ($\rW=0$), see Fig.~\ref{fig:IntPath}. Finally, Eq.~(\ref{B3}) ensures the asymptotic flatness of the metric (\ref{LE}), see (\ref{asflat}).

\subsection{Field equations and Linear Problem}
The stationary and axisymmetric vacuum Einstein equations for the metric potentials $U$ and $a$ are equivalent to the Ernst equation
\begin{equation}\label{Ernst}
 (\Re f)\Big(f_{,\rW\rW}+f_{,\zW\zW} +\frac{1}{\rW}f_{,\rW}\Big)
 = f_{,\rW}^2 + f_{,\zW}^2
\end{equation}
for the complex function
\begin{equation}\label{EP}
 f(\rW,\zW)=\ee^{2U(\rW,\zW)}+\ii b(\rW,\zW),
\end{equation}
where the twist potential $b$ is defined by
\begin{equation}\label{a}
 a_{,\rW} = \rW\ee^{-4U}b_{,\zW},\qquad
 a_{,\zW} = -\rW\ee^{-4U}b_{,\rW}.
\end{equation}
The potential $k$ can be calculated from
\begin{eqnarray}
 \label{k1}
 k_{,\rW} & = & \rW\Big[U_{,\rW}^2-U_{,\zW}^2+\frac{1}{4}\ee^{-4U}
         (b_{,\rW}^2-b_{,\zW}^2)\Big],\\
 \label{k2}
 k_{,\zW} & = & 2\rW\Big[U_{,\rW}U_{,\zW}+\frac{1}{4}\ee^{-4U}
         b_{,\rW}b_{,\zW}\Big].
\end{eqnarray}
As a consequence of the Ernst equation (\ref{Ernst}), the integrability conditions  $a_{,\rW\zW}=a_{,\zW\rW}$ and $k_{,\rW\zW}=k_{,\zW\rW}$ are satisfied such that the metric potentials $a$ and $k$ may be calculated via line integration from the Ernst potential~$f$.
 Since $\ee^{2U}=\Re f$, all metric coefficients in (\ref{LE}) can uniquely be determined from $f$.  Thus the boundary problem for the vacuum Einstein equations reduces to a boundary problem for the Ernst equation. However, we have to cope with non-local boundary conditions \eref{B1}-\eref{B3}, \eref{a}-\eref{k2} for the Ernst potential $f$.
Fortunately, these boundary conditions are well-adapted to the ``inverse method'', which we applied to solve the boundary value problem.

The Ernst equation is the integrability condition $\bPhi_{,z\bar z} = \bPhi_{,\bar zz}$ of the \emph{Linear Problem} (LP)~\cite{Neugebauer1980b,NeugebauerMeinel2003}
\begin{equation}\label{LP}
  \bPhi_{,z} = \left[\left(\begin{array}{cc}
                   N & 0\\
                   0 & M\end{array}\right)
                   +\lambda\left(\begin{array}{cc}
                   0 & N\\
                   M & 0\end{array}\right)\right]\bPhi,\\
           \end{equation}
       
\begin{equation}
\bPhi_{,\bar z} =\left[\left(\begin{array}{cc}
                   \bar M & 0\\
                   0 & \bar N\end{array}\right)
                   +\frac{1}{\lambda}\left(\begin{array}{cc}
                   0 & \bar M\\
                  \bar N & 0\end{array}\right)\right]\bPhi,
           \label{LPG}
        \end{equation}
where the \emph{pseudopotential}
$\bPhi(z,\bar z,\lambda)$ is a $2\times2$ matrix depending on the spectral parameter
\begin{equation}\label{lambda}
\lambda=\sqrt{\frac{K-\ii\bar z}{K+\ii z}}
\end{equation}
as well as on the complex coordinates 

\begin{equation}
 z=\rW+\ii\zW,\quad \bar z=\rW-\ii\zW,
\end{equation}
whereas $M$, $N$ and the complex conjugate quantities $\bar M$, $\bar N$ are functions of $z$, $\bar z$ (or $\rW$, $\zW$) alone and do not depend on the constant complex parameter $K$. 
Since the integrability condition must hold identical in $\lambda$ (or $K$) it yields a system of first order differential equations for 
$N$ and $M$ which is equivalent to the Ernst equation. The first order system has the ``first integrals''
\begin{equation}\label{firstin}
 M=\frac{f_{,z}}{f+\bar f},\qquad
 N=\frac{\bar f_{,z}}{f+\bar f}.
\end{equation}
Vice versa, any solution $f$ of the Ernst equation admits the unique determination
of the pseudopotential $\bPhi$ up to constants of integration. Thus the Ernst equation \eref{Ernst} and the LP \eref{LP}, \eref{LPG} are equivalent to each other.

Multiplying \eref{LP} by $\dd z$ and adding \eref{LPG} multiplied by $\dd{\bar z}$ one obtains the reformulation $\dd\bPhi = (...)\bPhi$
of the LP in the form of a system of (overdetermined) total differential equations.

 Without loss of generality we choose the standard representation
 \begin{equation}\label{norm}
 \bPhi=\left(\begin{array}{cc}
              \psi(\rW,\zW,\lambda) & \psi(\rW,\zW,-\lambda)\\
              \chi(\rW,\zW,\lambda) & -\chi(\rW,\zW,-\lambda)
             \end{array}\right)
\end{equation}
where
\begin{equation}\label{psibar}
 \bar \psi\left(\rW,\zW,\frac{1}{\bar\lambda}\right)=\chi(\rW,\zW,\lambda)
\end{equation}
due to the special structure of the coefficient matrices of the LP.
For $K\to\infty$ and $\lambda\to -1$, the functions $\psi$, $\chi$ can be normalized by
\begin{equation}\label{normali}
 \psi(\rW,\zW,-1)=\chi(\rW,\zW,-1)=1.
\end{equation}
As a consequence of the LP, the Ernst potential and the gravitomagnetic potential can be read off from the pseudopotential $\bPhi$ at
$K\to\infty$ and $\lambda\to +1$,
\begin{equation}
f(\rW,\zW)=\chi(\rW,\zW,1)=\Phii_{21}(\rW,\zW,1),
\label{fLP}
\end{equation}
\begin{equation}
 a(\rW,\zW)=\ii\ee^{-2U}\left.\left(K^2\diff{}{K}[\chi(-\lambda)-\psi(-\lambda)]\right)\right|_{\scriptsize\begin{array}{c}\lambda=1\\ K\to\infty\end{array}}-C,
  \label{aLP}
\end{equation} 
 where $C$ is an arbitrary constant.
The idea of the inverse (scattering) method is to discuss $\bPhi$, for fixed but arbitrary values of $\rW$, $\zW$ ($z$, $\bar z$) as a holomorphic function of $\lambda$ or $K$. In the latter case, $\bPhi$ ``lives'' on the two sheets of the Riemann surface associated with 
\eref{lambda}. As this mapping depends on the parameters $\rW$, $\zW$, the position of the branch points $K_{\rm B}=\ii\bar z$,
$\bar K_{\rm B}=-\ii z$ and the branch cut between them changes with the coordinates.

\section{Non-existence proof}

\subsection{Integration of the Linear Problem}
In order to solve the boundary problem \eref{Ernst}, \eref{B1}-\eref{B3}, we shall integrate (``solve'') the LP 
along the dashed line in Fig.~\ref{fig:IntPath} which marks, in the $\rW$-$\zW$ plane, the boundary of the vacuum domain outside the horizons. Starting from and returning to any axis point, say $\rW=0$, $\zW\in\Ap$, we shall make use of the boundary conditions and finally arrive at a representation of the Ernst potential
on the axis of symmetry. It turns out that this representation is sufficient to express all black hole quantities (such as areas $A_1$, $A_2$  and angular momenta $J_1$, $J_2$ of the black holes) in terms of three independent real parameters (plus two additional scaling parameters) and to establish the equations of state of the black hole thermodynamics of the equilibrium configuration under discussion. Furthermore, the axis values $f(\zW)=f(\rW=0,\zW)$ 
fix the solution $f(\rW,\zW)$ of the Ernst equation uniquely \cite{HauserErnst1981}.

Since $\lambda(K)$ as defined in (\ref{lambda}) degenerates on the $\zW$-axis, $\lambda=\pm1$, the LP $\dd\bPhi = (...)\,\bPhi$ can easily be integrated.  For $\lambda={+1}$ one obtains                              
\begin{equation}
 \mathcal A^\pm,\An,\mathcal H^{(i)}:\quad\bPhi=\left(\begin{array}{cc}\bar f & 1\\ f & -1\end{array}\right)
               \bf L,\quad \bf L=
               \left(\begin{array}{cc}A(K) & B(K)\\ C(K) & D(K)
                     \end{array}\right).
  \label{phiI}
\end{equation}
The representation for $\lambda={-1}$ follows from (\ref{norm}) by exchanging column elements. Remarkably, the $\bPhi$-matrix separates. The first factor depends on the path of integration whereas $\bf L$ representing the ``integration constants'' is a function of the spectral parameter $K$ alone. There is a difference between the case of two extended horizons and that of one or two degenerate horizons. In the first case one can parametrize the dashed curve by the coordinate $\zW$ everywhere on the $\zW$-axis, i.e.\ $f$=$f(\rW=0,\zW)$=$f(\zW)$ on $\mathcal A^\pm,\An,\mathcal H^{(i)}$. This is clearly impossible if the
dashed curve runs around a point-like horizon. A path like this can be described by an infinitesimal semicircle which brings (local) polar coordinates $(R,\theta)$ into play \cite {Meinel2008}.  
Then we have $f=f(R\to 0, \theta)=f(\theta)$,  $\theta\in[0,\pi]$ in \eref{phiI}.

To exploit characteristic properties of the horizons such as (\ref{KH}) and (\ref{kappa}), it is helpful to introduce corotating frames of reference defined by 
\begin{equation}\label{transf}
 \rW'=\rW,\quad \zW'=\zW,\quad\varphi'=\varphi-\Omega t,\quad t'=t,
\end{equation}
where $\Omega=\Omega_1$, $\Omega_2$. This coordinate transformation induces transformations of the gravitational potentials in~(\ref{LE}) such that
\begin{equation}\label{tran1}
 \ee^{2U'}\equiv\ee^{2V} = \ee^{2U}[(1+\Omega a)^2-\Omega^2\rW^2\ee^{-4U}],
\end{equation}
\begin{equation}
 (1-\Omega a')\ee^{2U'}=(1+\Omega a)\ee^{2U},
 \label{tran2}
\end{equation}
where a prime denotes ``corotating'' quantities. To determine the corotating Ernst potential $f'$ one has to apply~(\ref{a}) to $a'$ and $\ee^{2U'}$. Finally, using the equations \eref{firstin} for $N',M'$, one obtains the corotating pseudopotential \cite{NeugebauerMeinel2003}
\begin{equation}\label{phitran}
 \bPhi'={\bf T}_{\Omega}\bPhi,
\end{equation}
where
\begin{eqnarray}\nonumber\
{\bf T}_{\Omega} &=& \left(\begin{array}{cc}
             1+\Omega a-\Omega\rW\ee^{-2U} & 0\\
             0 & 1+\Omega a+\Omega\rW\ee^{-2U}
            \end{array}\right)\\ \label{Ttran}
  &&\quad+\ii(K+\ii z)\Omega\ee^{-2U}
   \left(\begin{array}{cc} -1 & -\lambda\\ \lambda & 1\end{array}\right).
\end{eqnarray}

The validity of the Ernst equations of the non-rotating and corotating system at the points of intersection
$\mathcal A /\mathcal H$ (axis of symmetry/extended horizon) and $\mathcal A/\mathcal C$ (axis/circle at infinity), see Fig.~\ref{fig:IntPath}, implies that $\bPhi$ and $\bPhi'$ must be continuous there as well. By way of example let us consider the continuity of $\bPhi$ in \eref{phiI} and $\bPhi'$ in \eref{phitran} at the point $\rW=0$, $\zW=K_1$. It immediately leads to a connection of the horizon and axis values of the ``integration constants'' ${\bf L}^{(1)}$ and ${\bf L}^{+}$, cf.~\eref{phiI},
\begin{equation}
{\bf L}^{(1)}=\left({\bf 1}-\frac{{\bf F}_{1}}{2\ii\Omega^{(1)}(K-K_1)}\right){\bf L}^{+},
 \label{AH}
\end{equation}
where ${\bf F}_1$ is is a special case of ${\bf F}_i$ as needed later,
\begin{equation}
  {\bf F}_i:=\left(\begin{array}{cc} -f_i & 1\\ -f_i^2 & f_i\end{array}\right),\quad f_i=f(K_i),\quad i=1,\dots,4. 
 \label{F1}
\end{equation}
Note that ${\bar f}_1 = -f_1$ since $\ee^{2V}$ and $\ee^{2U}$ are continuous at the points of intersection and 
$\ee^{2V} = 0$ from the side of the horizon.
If one continues the interlinking procedure along the closed contour, one returns to the starting point with the result
 \begin{equation}
 {\bf L}^{+}\left(\begin{array}{cc} 0 & 1\\ 1 & 0\end{array}\right)({\bf L}^{+})^{-1}=\Rp,
 \label{finres}
\end{equation}
where
\begin{equation}
 \Rp:=\prod\limits_{i=1}^4\left({\bf 1}-(-1)^i\frac{{\bf F}_i}{2\ii\Omega^{(i)}(K-K_i)}\right)\left(\begin{array}{cc}0 & 1\\ 1 & 0\end{array}\right)
 \label{Rdef}
\end{equation}
with $\Omega^{(1)}=\Omega^{(2)}=\Omega_1$, $\Omega^{(3)}=\Omega^{(4)}=\Omega_2$.

\emph{Point-like} (degenerate) horizons can be involved without any difficulty by setting $K_1=K_2$ or/and $K_3=K_4$ in these equations~\cite{Meinel2008}.

We shall show that (\ref{finres}) with (\ref{Rdef}) as the result of the integration of the LP along the closed (dashed) contour in Fig.~\ref{fig:IntPath} yields the Ernst potential on the axis.

\subsection{Ernst potential on the axis} 
At the branch points  $K=K_{\rm B}=\ii\bar z$, $K=\bar K_{\rm B}=-\ii z$ of the Riemann $K$-surface, where $\lambda= 0$, $\lambda= \infty$, respectively, one finds from (\ref{norm}) $\Phii_{11}=\Phii_{12}$, $\Phii_{21}=-\Phii_{22}$,
\begin{equation}
K=K_{\rm B}:\quad \bPhi\left(\begin{array}{cc} 0 & 1\\ 1 & 0\end{array}\right)\bPhi^{-1}=
 \left(\begin{array}{cc}1 & 0\\ 0 & -1\end{array}\right).
 \label{BP}
\end{equation}
On the $\zW$-axis one has \emph{confluent} branch points, $K_{\rm B}={\bar K}_{\rm B}=\zW$. For this choice one obtains from \eref{BP}, \eref{phiI} and \eref{finres} in terms of the Ernst potential $f^+$ on $\Ap$
\begin{equation}
  \Rp(\zW) =\left(\begin{array}{cc}{\bar f}^+(\zW) & 1\\ f^+(\zW) & -1\end{array}\right)^{-1}
  \left(\begin{array}{cc} 1 & 0\\ 0 & -1\end{array}\right)
  \left(\begin{array}{cc} {\bar f}^+(\zW) & 1\\ f^+(\zW) & -1\end{array}\right)
 \label{Rf}
\end{equation}
and so
\begin{equation}
[\Rp(\zW)-{\bf 1}]\left(\begin{array}{c} 1\\ f^+(\zW)\end{array}\right)={\bf 0}
 \label{EW}
\end{equation}
for $\Rp(\zW):=\Rp(K=\zW)$, cf.~\eref{Rdef}.
Note that (\ref{EW}) is equivalent to (\ref{Rf}): a second column resulting from (\ref{Rf}) and (\ref{EW}) are complex conjugate. 

 We shall now discuss  properties of the Ernst potential on the axis which can be derived from the eigenvalue equation 
(\ref{EW}). First of all, let us point out that similar equations can be derived for all intervals $\mathcal A$,
$\mathcal H$ by the interlinking procedure as explained in (\ref{AH}).     
At the first glance, $f^+(\zW)$ seems to be a quotient of two polynomials of fourth degree in $\zW$. However, attention must be paid to the fact that the characteristic determinant has to vanish,
\begin{equation}
|\Rp(\zW)-{\bf 1}|={\bf 0}
 \label{detR}.
\end{equation}
This condition tells us that the numerator and the denominator of $f^+(\zW)$ must have two common zeros such that  
\emph{the axis potential is a quotient of two (normalized) polynomials of second degree in $\zW$},
\begin{equation}\label{axisf}
 f^+(\zW)=\frac{n_2(\zW)}{d_2(\zW)}=
 \frac{\zW^2+q\zW+r}{\zW^2+s\zW+t},
\end{equation}
where $q$, $r$, $s$, $t$ are complex constants which can be expressed in terms of $f_i$, $K_i$ and $\Omega^{(i)}$, $(i=1,\dots,4)$. For \emph{extended} (sub-extremal) horizons the following reparametrization is useful:

Defining 
\begin{equation}
 \alpha_i:=\frac{\bar d_2(K_i)}{d_2(K_i)},\quad \alpha_i{\bar\alpha}_i=1,\quad
 \beta_i:=\frac{\bar n_2(K_i)}{n_2(K_i)},\quad \beta_i{\bar\beta}_i=1 
 \label{alphi}
\end{equation}
and using
\begin{equation}
 \overline{f^+(K_i)}=-f^+(K_i), \qquad i=1,\dots,4,
\label{albe}
\end{equation}
one obtains
\begin{equation}\label{beal}
 \beta_i=-\alpha_i.
\end{equation}
The equations (\ref{alphi}) form a linear algebraic system for the parameters $q$, $r$, $s$, $t$. Eliminating them in (\ref{axisf}) one arrives
at a determinant representation for the \emph{Ernst potential on the axis} $\mathcal A^+$, $f^+(\zW)$, which can be written in the form 
\begin{equation}
 f^+(\zW) = \frac{\left|\begin{array}{cc}
              s_{12}-1 & s_{14}-1\\
              s_{23}-1 & s_{34}-1\end{array}\right|}
              {\left|\begin{array}{cc}
              s_{12}+1 & s_{14}+1\\
              s_{23}+1 & s_{34}+1\end{array}\right|},\qquad
 s_{ij}:=\frac{\alpha_i(\zW-K_i)-\alpha_j(\zW-K_j)}{K_{ij}},
 \label{Yax}
\end{equation}
where
\begin{equation}
 K_{ij}:=K_i-K_j,\qquad i,j=1,\dots,4.
\end{equation}
The continuation of $f^+(\zW)$ to all space is unique~\cite{HauserErnst1981} and leads to the representation 
\begin{equation}
 f(\rW,\zW) = \frac{\left|\begin{array}{cc}
              R_{12}-1 & R_{14}-1\\
              R_{23}-1 & R_{34}-1\end{array}\right|}
              {\left|\begin{array}{cc}
              R_{12}+1 & R_{14}+1\\
              R_{23}+1 & R_{34}+1\end{array}\right|},\qquad
 R_{ij}:=\frac{\alpha_ir_i-\alpha_jr_j}{K_{ij}},
 \label{Yamazaki}
\end{equation}
where
\begin{equation}
 r_i := \sqrt{(\zW-K_i)^2+\rW^2}\ge0,\qquad i,j=1,\dots,4.
 \label{ri}
\end{equation}
$f(\rW,\zW)$ is the Ernst potential of the \emph{double-Kerr-NUT solution} which was originally generated by a two-fold B\"acklund transformation of Minkowski space \cite{Neugebauer1980,KramerNeugebauer1980} in the form of a quotient of two $5\times5$ determinants. According to a rule of Yamazaki \cite{Yamazaki1983}, this type of determinants can be reduced to $2\times 2$ determinants as used in \eref{Yamazaki}. Making use of \eref{a} [or \eref{aLP}], \eref{k1}, \eref{k2} and $\ee^{2U}=\Re f$, one finds determinant representations for all metric coefficients, i.e.\ for $a$, $k$, $\ee^{2U}$ in \eref{LE}, see~\cite{NeugebauerHennig2012}.

 As a condition identical in $K$, Eq.~\eref{detR} yields four constraints among the parameters $\Omega_1$, $\Omega_2$; 
$K_1-K_2$, $K_2-K_3$, $K_3-K_4$; $f_1$, $\dots$, $f_4$. It can be shown \cite{NeugebauerHennig2012} that this system of algebraic equations guarantees that $a=0$ on $\mathcal A^\pm,\An$ and $\mathcal C$. As a consequence, it eliminates NUT parameters from the Ernst potential. In consideration of (\ref{EW}) and introducing dimensionless coordinates
$\tilde\rW$, $\tilde\zW$ via
\begin{equation}
 \tilde\rW = \frac{\rW}{K_{23}},\qquad
 \tilde\zW = \frac{\zW-K_1}{K_{23}}
 \label{dles}
\end{equation}   
one realizes that $f^+(\zW)$ from (\ref{EW}) and therefore $f(\rW,\zW)$ depend on four real parameters.
 
\subsection{ Weyl-Bach force between the black holes}
So far we have examined the Ernst equation as a classical field equation. It is questionable whether the parameter conditions 
(\ref{detR}) alone could rule out the Ernst potential under discussion and lead to a non-existence proof. 
Consider the static Ernst equation which is an axisymmetric Laplace equation for the ``Newtonian'' gravitational potential $U$
in Weyl coordinates. The superposition of two  solutions with aligned rod-shaped sources (as  ``classical'' precursors of horizons) is regular outside the sources. It is the gravitational interaction (``force'') between the rods that forbids equilibrium. Bach, who examined this example in the already mentioned Bach/Weyl paper~\cite{Bach1922}, noted that the metric function $k$ cannot vanish on the portion of the axis between the two sources and that this fact violates the regularity
of the solution. Weyl's remarks (published as a supplement to Bach's paper) focused on the interpretation of this type of
singularity. He used fictitious stresses described by an energy-momentum tensor to define a force of attraction between the
sources. This ``Weyl-Bach force'' turned out to be proportional to a constant value of $k$ on the portion of the axis between the two sources (in our notation $k^0$) provided that $k^+=0$, $k^-=0$, which is a possible gauge. Note that $k$ as defined in \eref{k1}, \eref{k2} has  constant values on \emph{all} intervals of  the $\zW$-axis. One of them, considered to be the arbitrary integration constant, can be chosen so that, say, $k^+=0$. Integration of \eref{k1}, \eref{k2}  along $\mathcal C$ in~Fig.~\ref{fig:IntPath} results in $k^-=0$.

 Equipped with that physical as well as  geometrical interpretation of the superpotential $k^0$ (``attractive force'' that violates the Lorentzian geometry in the vicinity of the rotation axis) we shall examine the boundary condition 
\begin{equation}
k^0=0
 \label{force}
\end{equation}
for the \emph{gauge}
\begin{equation}
k^+=k^-=0.
 \label{gaug}
\end{equation}
Our discussion is based on the original parametrization of the double Kerr-NUT solution, see \eref{alphi}, \eref{beal}, \eref{Yax}, and Kramer's representation of $\ee^{2k}$ \cite{Kramer1986}. (In principle, one could determine  the axis values of this gravitational potential from the axis values of the Ernst potential by integrating the equations \eref{k1}, \eref{k2} along the axis, i.e.\ by operations on the dashed contour in Fig.~\ref{fig:IntPath}.) It turns out that the boundary conditions for non-overlapping extended horizons \eref{force}, \eref{gaug} are satisfied by only one parameter condition:
\begin{equation}
 \alpha_1\alpha_2+\alpha_3\alpha_4=0.
 \label{Bed1}
\end{equation}
Two of the four conditions (\ref{detR}) can be used to eliminate $\Omega_1$ and $\Omega_2$. The two remaining equations turn out to be equivalent to the equations 
\begin{equation}
  \frac{(1-\alpha_4)^2}{\alpha_4}w^2= \frac{(1-\alpha_3)^2}{\alpha_3},\quad w:=\sqrt{\frac{K_{14}K_{24}}{K_{13}K_{23}}}\in[1,\infty),
 \label{w1}
\end{equation}

\begin{equation}
 \frac{(1+\alpha_2)^2}{\alpha_2}w'^2=\frac{(1+\alpha_1)^2}{\alpha_1},\quad w':=\sqrt{\frac{K_{23}K_{24}}{K_{13}K_{14}}}\in(0,1].
 \label{w2}
\end{equation}
The three restrictions \eref{Bed1}, \eref{w1}, \eref{w2} are nothing else but a reformulation of the original Tomimatsu-Kihara conditions~\cite{Kihara1982,TomimatsuKihara1982}. This reformulation is due to essential examinations of the double-Kerr-NUT solution by 
Manko, Ruiz and Sanabria-G\'omez~\cite{Manko2000,Manko2001}.
As was particularly shown in~\cite{Manko2001}, the restrictions \eref{Bed1}, \eref{w1}, \eref{w2} can be solved to express the parameters $\alpha_1,\dots,\alpha_4$ in terms of the three real parameters $w$, $w'$ and $\phi$:
\begin{equation}
  \alpha_1 = \frac{w'\alpha^2+\ii\eps\alpha}{w'-\ii\eps\alpha},\quad
  \alpha_2 = \frac{\alpha^2+\ii w'\eps\alpha}{1-\ii w'\eps\alpha},
 \label{alf1}
\end{equation}

\begin{equation}
 \alpha_3 = \frac{w\alpha^2-\alpha}{w-\alpha},\quad
 \alpha_4 = \frac{\alpha^2-w\alpha}{1-w\alpha},
 \label{alf2}
\end{equation}
where $\eps=\pm 1$ and 
\begin{equation}
 \alpha_3\alpha_4 = -\alpha_1\alpha_2 \equiv \alpha^2\qquad
 (\alpha=\ee^{\ii\phi},~\phi \in[0,2\pi)).
\end{equation}
Now we have  arrived at the final form of the solution of the boundary problem \eref{B1}, \eref{B2}, \eref{B3} for the Ernst equation \eref{Ernst}. Eliminating the $\alpha_i$ in favour of $w$, $w'$, $\phi$ and introducing dimensionless
coordinates (\ref{dles}), $f$ becomes a function of two coordinates and three real parameters\footnote{From this point of view, the Ernst potential of the Kerr solution depends on one real parameter and two dimensionless coordinates.},   
\begin{equation}
f=f(\tilde\rW,\tilde\zW;w,w',\phi).
 \label{P3}
\end{equation}
Note that the relative horizon ``lengths'' $K_{12}/K_{23}$, $K_{34}/K_{23}$ can be expressed in terms of $w, w'$ as well.
At this point we cannot guarantee that this Ernst potential is well-behaved. Computer experiments show that the regularity
of the Ernst potential on the axis of symmetry must be ``paid of'' in the form of singular rings outside the horizons.
\begin{figure}
 \centering
 \includegraphics[scale=0.8]{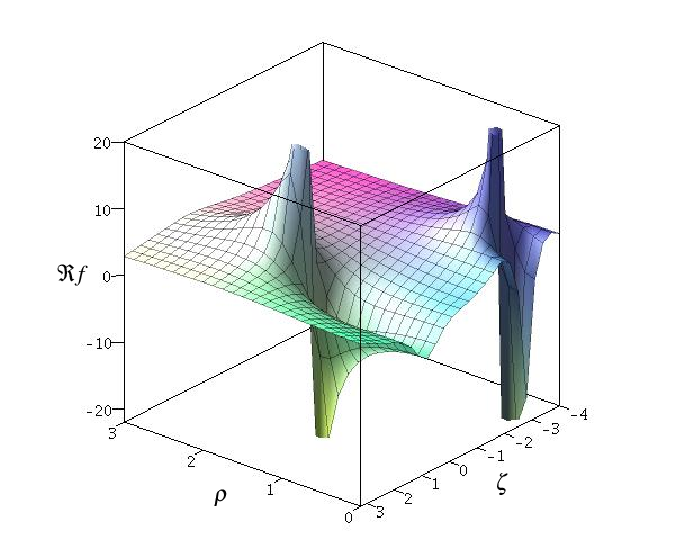}
 \caption{The singular Ernst potential for a particular example configuration. Parameters: $\phi=-0.1$, $w=1.3$, $w'=0.5$, $K_1=2$, $K_{23}=2$, $\eps=1$.}
 \label{fig:singular}
\end{figure}
Fig.~\ref{fig:singular} conveys an impression of the structure of this type of singularity. Irregularities on the horizons could also rule out
the solution. Since area $A$ and angular momentum $J$ are characteristic parameters for each axisymmetric black hole, it was obvious to examine restrictions of these parameters. Based on the literature as commented on in the introduction, we could take for granted that the inequality $8\pi|J| < A$ has to hold on both horizons of a regular spacetime with two sub-extremal black holes.

\subsection{Angular momentum-area inequality and non-existence proof}
In order to examine the inequalities
\begin{equation}
\mathcal H^{(i)}:\quad 8\pi|J_i|<A_i,\quad i=1,2,
 \label{ineq}
\end{equation}
we calculate the quantities
\begin{equation}
p_i:=\frac{8\pi J_i}{A_i},\qquad i=1,2,
\end{equation}
in terms of the parameters $w$, $w'$ and $\phi\in[0,2\pi)$. Note that this computation can be performed with the aid of the axis values of the Ernst potential alone, see \cite{NeugebauerHennig2009}. The result is
\begin{equation}\label{pJ}
 p_1 = \eps\frac{1+\Phi w'}{w'(\Phi+w')},\quad
 p_2 = \eps\frac{w(w-\Phi)}{1-w\Phi}
\end{equation}
where
\begin{equation}
 \Phi:=\cos\phi+\eps\sin\phi,\quad \eps=\pm1.
\end{equation}
From this we have
\begin{equation}
  p_1^2-1= (1-w'^2)\frac{w'^2+2\Phi w'+1}{w'^2(\Phi+w')^2}<0
\end{equation}
and
\begin{equation}
  p_2^2-1= (w^2-1)\frac{w^2-2\Phi w+1}{(w\Phi-1)^2}<0.
\end{equation}
For the allowed parameter ranges $w\in[1,\infty)$,
$w'\in(0,1]$, which follow from the definitions \eref{w1}, \eref{w2} of $w$ and $w'$, these inequalities can only hold if
\begin{equation}
w'^2+2\Phi w'+1<0\qquad\textrm{and}\qquad
w^2-2\Phi w+1<0.
\end{equation}
This, however, implies $\Phi w'<0$ and $\Phi w>0$ in contradiction to $w'>0$ and $w>0$.
Thus \emph{we have proved the non-existence of stationary and axisymmetric
configurations consisting of two aligned sub-extremal black holes} and we conclude that the spin-spin repulsion cannot compensate for the gravitational attraction.

\section{Summary\label{sec:summary}}

\begin{figure}
 \centering
 \includegraphics[scale=0.75]{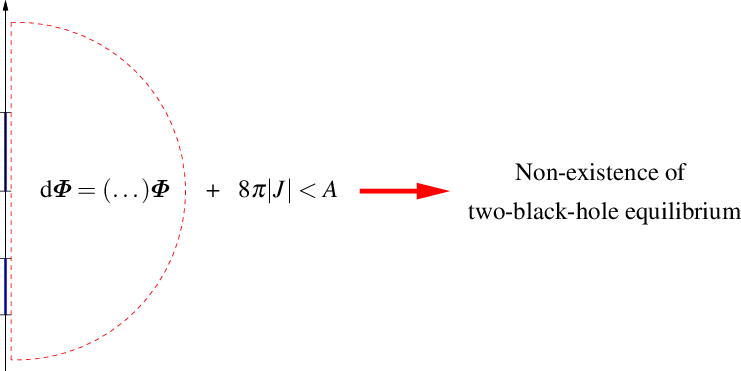}
 \caption{Summary of the non-existence proof for two sub-extremal black holes.}
 \label{fig:sum}
\end{figure}

As a characteristic example for the ongoing discussion about existence or non-existence of stationary equilibrium configurations within the theory of General Relativity, we have studied the question whether two aligned, sub-extremal black holes can be in equilibrium. The result of our above analysis, whose details can be found in \cite{NeugebauerHennig2009,HennigNeugebauer2011,NeugebauerHennig2012}, is negative: \emph{there are no two-black-hole equilibrium configurations!} 
The idea of the non-existence proof is illustrated in Fig.~\ref{fig:sum} and can be summarized as follows. 

Equilibrium configurations with two aligned rotating black holes, if any existed, can be described by a boundary value problem for two separate \mbox{(Killing-)} horizons. Remarkably, this problem can be solved 
by integrating the Linear Problem $\dd\bPhi=(\dots)\bPhi$ along the dashed contour as sketched in Fig.~\ref{fig:IntPath} or Fig.~\ref{fig:sum}. Thus we arrive necessarily at particular Kerr-NUT solutions which have two horizons [at least according to the definitions \eref{KH} and \eref{kappa}] and show the 
correct regular behaviour at infinity and on the symmetry axis, whereas regularity off the axis is not guaranteed. 
On the contrary, we find that all candidate solutions indeed do suffer from irregularities. One of them is the violation of the angular momentum-area inequality $8\pi |J|<A$, which must hold for any regular sub-extremal black hole. We could show that there is no choice of parameters for which angular momentum and area of the two horizons jointly satisfy the inequality. Hence, there exists no regular solution of the vacuum field equations for  stationary  two-black-hole configurations.
For brief comments on the extension of the non-existence proof to \emph{degenerate} black holes, see the following supplement and Fig.~\ref{fig:BVPs}a. 

\section{Supplement}

\subsection{Degenerate horizons}

\begin{figure}
 \centering
 \includegraphics[scale=0.7]{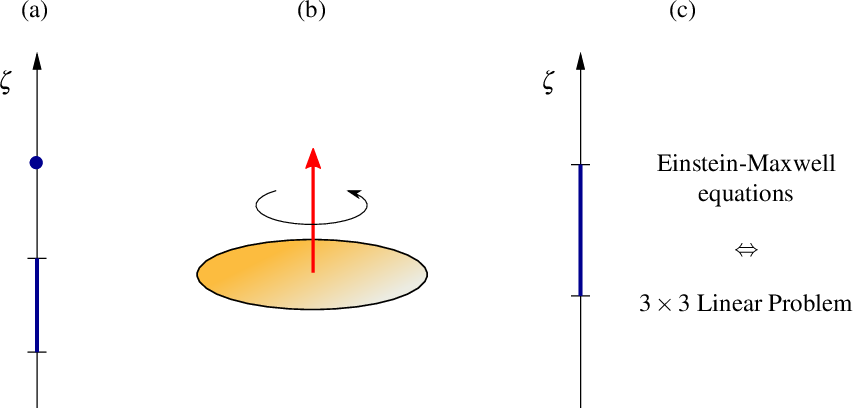}
 \caption{Illustration of further BVPs that can be solved with the inverse scattering method: (a) two-horizon problem with one degenerate black hole, (b) rigidly rotating disk of dust, (c) constructive uniqueness proof for the Kerr-Newman black hole.}
 \label{fig:BVPs}
\end{figure}

The analysis as presented above only applies to configurations with two \emph{sub-extremal (``extended'') horizons}. But we have already indicated several times that it is possible to extend the proof to configurations containing one or even two \emph{degenerate black holes} (with ``point-like'' horizons, where $K_1=K_2$ or/and $K_3=K_4$), see Fig.~\ref{fig:BVPs}a. In the first part of this supplement we give an outline of this generalisation. For details we refer to \cite{HennigNeugebauer2011}.

As already mentioned, the degeneracy of a Killing horizon to a point is merely a peculiarity of the Weyl-Lewis-Papapetrou coordinates. In order to resolve the internal structure of the horizon, we have discussed the LP in polar coordinates centred at the point-like horizon. Integrating the LP, we found that all possible equilibrium configurations with degenerate black holes can be obtained as particular limits of the double-Kerr-NUT family of solutions. It turned out that there are two families of (two-parametric) solutions describing configurations with one degenerate and one extended horizon and  three (one-parametric) solution families for configurations with two degenerate horizons. In order to exclude these families as acceptable equilibrium configurations, we showed that they suffer from unphysical singularities.

In the case of one degenerate and one sub-extremal black hole, the angular momentum-area inequality becomes an \emph{equality} for the degenerate horizon (which turned out to be satisfied identically and, therefore, did not provide any new information). The inequality for the sub-extremal black hole restricts the parameters, but does not yet exclude the possibility of a regular equilibrium configuration. Hence an additional ingredient was required for the desired non-existence proof, which was the positivity of the total mass (ADM mass) of the spacetime. As first shown by Schoen and Yau \cite{SchoenYau1979,SchoenYau1981}, the mass of a regular, asymptotically flat spacetime satisfying the dominant energy condition (which is certainly satisfied for a black hole vacuum spacetime) is positive. However, for the configuration in question, we found that the entire parameter range in which the sub-extremal black hole satisfies the angular momentum-area inequality has a \emph{
negative} mass. Thus we could conclude that the solutions with one degenerate and one sub-extremal horizon are singular.

Finally, we studied the three solution branches for configurations with two degenerate horizons. One branch could be excluded since it has a negative ADM mass for all possible parameter values. The other two branches have negative masses for most parameter values, but there are small parameter regions with positive mass. Fortunately, these solution branches are relatively simple and, by studying the solutions of a certain quartic equation, it was possible to demonstrate explicitly that singularities (singular rings around the $\zW$-axis) are present even in the parameter range with positive ADM mass. 

Hence  we could extend the non-existence proof to \emph{all} forms of horizons.

\subsection{Further applications of the inverse method}

In the following we briefly comment on some other applications of the inverse (scattering) method to rotating objects. The integration of the Linear Problem of the Ernst equation  $\dd\bPhi=(\dots)\bPhi$ along a suitable closed contour was first practised to determine the gravitational field of a rigidly rotating disk of dust \cite{NeugebauerMeinel1993, NeugebauerMeinel1994, NeugebauerMeinel1995}, see Fig.~\ref{fig:BVPs}b. Among other things, the solution (expressed in terms of theta functions) describes a parametric collapse of the disk with a final phase transition to an extreme black hole. 

Obviously, the integration method under discussion can be used to construct the Kerr solution as the unique solution of the one-horizon boundary problem \cite{Neugebauer2000}. This corresponds to the methods of electrostatics and is an alternative to ``complex tricks'' and other formal derivations of the Kerr metric in the textbooks.

Extending the vacuum examinations to Einstein-Maxwell fields, Meinel \cite{Meinel2012} has recently constructed the Kerr-Newman solution by integrating a $3\times3$ electrovacuum LP, see Fig.~\ref{fig:BVPs}c and thus proved the uniqueness of the solution.

\section*{Acknowledgements}
We would like to thank Reinhard Meinel, Marcus Ansorg and Andreas Kleinw\"achter for many valuable discussions and Ben Whale for commenting on the manuscript.

\section*{References}
\bibliography{ae100prgNeugebauerHennig}

\begin{thebibliography}{10}

\bibitem{AnsorgPetroff2006}
Ansorg, M.  and Petroff, D., ``Negative Komar mass of single objects in
  regular, asymptotically flat spacetimes'', {\em Class. Quantum Grav.}, {\bf
  23}, L81, (2006).

\bibitem{AnsorgPfister2008}
Ansorg, M.  and Pfister, H., ``A universal constraint between charge and
  rotation rate for degenerate black holes surrounded by matter'', {\em Class.
  Quantum Grav.}, {\bf 25}, 035009, (2008).

\bibitem{Bach1922}
Bach, R.  and Weyl, H., ``Neue L\"osungen der Einsteinschen
  Gravitationsglei\-chungen'', {\em Mathemat. Z.}, {\bf 13}, 134, (1922).
  Republication in English: \emph{Gen. Relativ. Gravit.}, {\bf 44}, 817,
  (2012).

\bibitem{BelinskiZakharov1978}
Belinski\u{i}, V.~A.  and Zakharov, V.~E., ``Integration of the Einstein
  equations by means of the inverse scattering problem technique and
  construction of exact soliton solutions'', {\em Pis'ma Zh. Eksp. Teor. Fiz.}, {\bf
  75}, 1955, (1978). (in Russian) English translation: \emph{Sov. Phys. JETP},
  {\bf 48}, 985 (1978).

\bibitem{BelinskiZakharov1979}
Belinski\u{i}, V.~A.  and Zakharov, V.~E., ``Stationary gravitational solitons
  with axial symmetry'', {\em Pis'ma Zh. Eksp. Teor. Fiz.}, {\bf 77}, 3, (1979). (in
  Russian) English translation: \emph{Sov. Phys. JETP}, {\bf 50}, 1 (1979).

\bibitem{BoothFairhurst2008}
Booth, I.  and Fairhurst, S., ``Extremality conditions for isolated and
  dynamical horizons'', {\em Phys. Rev. D}, {\bf 77}, 084005, (2008).

\bibitem{Carter1973}
Carter, B., ``Black hole equilibrium states'', in deWitt, C.  and deWitt, B.,
  eds., {\em Black holes (Les Houches)}. Gordon and Breach, London, (1973).

\bibitem{Chrusciel2011}
Chru\'{s}ciel, P.~T., Eckstein, M., Nguyen, L.  and Szybka, S.~J., ``Existence
  of singularities in two-Kerr black holes'', {\em Class. Quantum Grav.}, {\bf
  28}, 245017, (2011).

\bibitem{Dain2011}
Dain, S., ``Geometric inequalities for axially symmetric black holes'',
  submitted, (2011). arXiv:1111.3615 [gr-qc].

\bibitem{DainReiris2011}
Dain, S.  and Reiris, M., ``Area-angular momentum inequality for axisymmetric
  black holes'', {\em Phys. Rev. Lett.}, {\bf 107}, 051101, (2011).

\bibitem{DietzHoenselaers1985}
Dietz, W.  and Hoenselaers, C., ``Two mass solution of Einstein's vacuum
  equations: the double Kerr solution'', {\em Ann. Phys.}, {\bf 165}, 319,
  (1985).

\bibitem{HauserErnst1981}
Hauser, I.  and Ernst, F.~J., ``Proof of a Geroch conjecture'', {\em J. Math.
  Phys.}, {\bf 22}, 1051, (1981).

\bibitem{Hennig2008}
Hennig, J., Ansorg, M.  and Cederbaum, C., ``A universal inequality between the
  angular momentum and horizon area for axisymmetric and stationary black holes
  with surrounding matter'', {\em Class. Quantum Grav.}, {\bf 25}, 162002,
  (2008).

\bibitem{HennigNeugebauer2011}
Hennig, J.  and Neugebauer, G., ``Non-existence of stationary two-black-hole
  configurations: the degenerate case'', {\em Gen. Relativ. Gravit.}, {\bf 43},
  3139, (2011).

\bibitem{Hoenselaers1984}
Hoenselaers, C., ``Remarks on the Double-Kerr-Solution'', {\em Prog. Theor.
  Phys.}, {\bf 72}, 761, (1984).

\bibitem{HoenselaersDietz1983}
Hoenselaers, C.  and Dietz, W., Talk given at the GR10 meeting, Padova, (1983).

\bibitem{Kihara1982}
Kihara, M.  and Tomimatsu, A., ``Some properties of the symmetry axis in a
  superposition of two Kerr solutions'', {\em Prog. Theor. Phys.}, {\bf 67},
  349, (1982).

\bibitem{Kramer1986}
Kramer, D., ``Two Kerr-NUT constituents in equilibrium'', {\em Gen. Relativ.
  Gravit.}, {\bf 18}, 497, (1980).

\bibitem{KramerNeugebauer1980}
Kramer, D.  and Neugebauer, G., ``The superposition of two Kerr solutions'',
  {\em Phys. Lett. A}, {\bf 75}, 259, (1980).

\bibitem{Krenzer2000}
Krenzer, G., ``Schwarze L\"ocher als Randwertprobleme der
  axialsymmetrisch-station\"aren Einstein-Gleichungen'', PhD Thesis, University
  of Jena, (2000).

\bibitem{Manko2001}
Manko, V.~S.  and Ruiz, E., ``Exact solution of the double-Kerr equilibrium
  problem'', {\em Class. Quantum Grav.}, {\bf 18}, L11, (2001).

\bibitem{Manko2000}
Manko, V.~S., Ruiz, E.  and Sanabria-G\'omez, J.~D., ``Extended multi-soliton
  solutions of the Einstein field equations: II. Two comments on the existence
  of equilibrium states'', {\em Class. Quantum Grav.}, {\bf 17}, 3881, (2000).

\bibitem{Meinel2012}
Meinel, R., ``Constructive proof of the Kerr-Newman black hole uniqueness
  including the extreme case'', {\em Class. Quantum Grav.}, {\bf 29}, 035004,
  (2012).

\bibitem{Meinel2008}
Meinel, R., Ansorg, M., Kleinw\"achter, A., Neugebauer, G.  and Petroff, D.,
  {\em Relativistic figures of equilibrium}, (Cambridge University Press,
  Cambridge, 2008).

\bibitem{Neugebauer1980}
Neugebauer, G., ``A general integral of the axially symmetric stationary
  Einstein equations'', {\em J. Phys. A}, {\bf 13}, L19, (1980).

\bibitem{Neugebauer1980b}
Neugebauer, G., ``Recursive calculation of axially symmetric stationary
  Einstein fields'', {\em J. Phys. A}, {\bf 13}, 1737, (1980).

\bibitem{Neugebauer1996}
Neugebauer, G., ``Gravitostatics and rotating bodies'', in Hall, G.~S.  and
  Pulham, J.~R., eds., {\em Proc. 46th Scottish Universities Summer School in
  Physics (Aberdeen)}. Copublished by SUSSP Publications, Edinburgh, and
  Institute of Physics Publishing, London, (1996).

\bibitem{Neugebauer2000}
Neugebauer, G., ``Rotating bodies as a boundary value problems'', {\em Ann.
  Phys. (Leipzig)}, {\bf 9}, 342, (2000).

\bibitem{NeugebauerHennig2009}
Neugebauer, G.  and Hennig, J., ``Non-existence of stationary two-black-hole
  configurations'', {\em Gen. Relativ. Gravit.}, {\bf 41}, 2113, (2009).

\bibitem{NeugebauerHennig2012}
Neugebauer, G.  and Hennig, J., ``Stationary two-black-hole configurations: A
  non-existence proof'', {\em J. Geom. Phys.}, {\bf 62}, 613, (2012).

\bibitem{NeugebauerMeinel1993}
Neugebauer, G.  and Meinel, R., ``The Einsteinian gravitational field of a
  rigidly rotating disk of dust'', {\em Astrophys. J.}, {\bf 414}, L97, (1993).

\bibitem{NeugebauerMeinel1994}
Neugebauer, G.  and Meinel, R., ``General relativistic gravitational field of a
  rigidly rotating disk of dust: Axis potential, disk metric and surface mass
  density'', {\em Phys. Rev. Lett.}, {\bf 73}, 2166, (1994).

\bibitem{NeugebauerMeinel1995}
Neugebauer, G.  and Meinel, R., ``General relativistic gravitational field of a
  rigidly rotating disk of dust: Solution in terms of ultraelliptic
  functions'', {\em Phys. Rev. Lett.}, {\bf 75}, 3046, (1995).

\bibitem{NeugebauerMeinel2003}
Neugebauer, G.  and Meinel, R., ``Progress in relativistic gravitational theory
  using the inverse scattering method'', {\em J. Math. Phys.}, {\bf 44}, 3407,
  (2003).

\bibitem{SchoenYau1979}
Schoen, R.  and Yau, S.-T., ``On the positive mass conjecture in general
  relativity'', {\em Commun. Math. Phys.}, {\bf 65}, 45, (1979).

\bibitem{SchoenYau1981}
Schoen, R.  and Yau, S.-T., ``Proof of the positive mass theorem. II'', {\em
  Commun. Math. Phys.}, {\bf 79}, 231, (1981).

\bibitem{TomimatsuKihara1982}
Tomimatsu, A.  and Kihara, M., ``Conditions for regularity on the symmetry axis
  in a superposition of two Kerr-NUT solutions'', {\em Prog. Theor. Phys.},
  {\bf 67}, 1406, (1982).

\bibitem{Varzugin1997}
Varzugin, G., ``Equilibrium configuration of black holes and the inverse
  scattering method'', {\em Theoret. Math. Phys.}, {\bf 111}, 667, (1997).

\bibitem{Varzugin1998}
Varzugin, G., ``The interaction force between rotating black holes in
  equilibrium'', {\em Theoret. Math. Phys.}, {\bf 116}, 1024, (1998).

\bibitem{Yamazaki1983}
Yamazaki, M., ``Stationary line of $N$ Kerr masses kept apart by gravitational
  spin-spin interaction'', {\em Phys. Rev. Lett.}, {\bf 50}, 1027, (1983).

\end{thebibliography}

\end{document}